\documentclass[aps,showpacs,twocolumn,superscriptaddress]{revtex4}

\usepackage{graphicx}  
\usepackage{dcolumn}
\usepackage{amsmath}

\newcommand{\vp}{{\vec p}}

\newcommand{\ds}{\displaystyle}

\begin{document}
%
\preprint{ANL-PHY-9815-TH-2001, UNITU-THEP-01/2001}

\title{Plasma production and thermalisation in a strong field}


\author{D.V. Vinnik} 
\affiliation{Institut f\"ur Theoretische Physik, Universit\"at T\"ubingen,
Auf der Morgenstelle 14, D-72076 T\"ubingen, Germany\vspace*{1ex}}
%
%
\author{A.V.~Prozorkevich}
\author{S.A.~Smolyansky} 
\affiliation{Physics Department, Saratov State University, 410071 Saratov,
Russian Federation\vspace*{1ex}}
\author{V.D.~Toneev}
\affiliation{Bogoliubov Laboratory of Theoretical Physics,\\ Joint Institute
for Nuclear Research, 141980 Dubna, Russian Federation\vspace*{1ex}}
\author{M.B.~Hecht}
\author{C.D.~Roberts}  
\affiliation{Physics Division, Bldg 203, Argonne National
Laboratory, Argonne Illinois 60439-4843\vspace*{1ex}}
\author{S.M.~Schmidt\vspace*{1ex}}
\affiliation{Institut f\"ur Theoretische Physik, Universit\"at T\"ubingen, Auf
der Morgenstelle 14, D-72076 T\"ubingen, Germany\vspace*{1ex}}

\date{\today\vspace*{1ex}}
\begin{abstract}
\rule{0ex}{3ex} 
Aspects of the formation and equilibration of a quark-gluon plasma are
explored using a quantum kinetic equation, which involves a non-Markovian,
Abelian source term for quark and antiquark production and, for the collision
term, a relaxation time approximation that defines a time-dependent
quasi-equilibrium temperature and collective velocity.  The strong Abelian
field is determined via the simultaneous solution of Maxwell's equation.  A
particular feature of this approach is the appearance of plasma oscillations
in all thermodynamic observables.  Their presence can lead to a sharp
increase in the time-integrated dilepton yield.
\end{abstract}
%
\maketitle

\section{Introduction}
The Relativistic Heavy-Ion Collider (RHIC) at Brook\-haven National
Laboratory and the Large Hadron Collider (LHC) at CERN are designed with the
goal of producing an equilibrated phase of deconfined partonic matter: the
quark gluon plasma (QGP).  Lattice-QCD simulations~\cite{edwin} and
well-constrained phenomenological models~\cite{bastirev} predict a second
order phase transition at $T_c \simeq 150\,$MeV; i.e., at energy densities
$\gtrsim 1\,$GeV/fm$^3$, in the equilibrium two light-flavour theory.  For
more than two flavours the characteristics of the transition are not as
clear, but there is a transition.  A number of phenomena have been proposed
as signals for the existence of an equilibrated QGP~\cite{signals} but the
nonequilibrium stages in the plasma's development are poorly understood and
it is a contemporary challenge to develop a description of the spacetime
evolution of an ultra-relativistic heavy-ion collision (URHIC): from particle
production in the collision, through equilibration and plasma formation, and
on to hadronisation.

Two methods are commonly used to describe the production of partons in a
collision: a perturbative pre-formed parton picture~\cite{Gribovpartons} and
a nonperturbative flux-tube-based picture~\cite{anders83}.  They are
complementary, and Monte-Carlo event
generators~\cite{partonMC,Geiger95,stringMC} and hydrodynamical
models~\cite{hydro} have been developed to facilitate the analysis of data
using either production model.

Herein we focus on the pre-equilibrium particle-production stage in the
evolution of a QGP and choose to employ a flux-tube model.  In this model the
two colliding nuclei are imagined to pass through one another and stretch a
high energy-density flux-tube between themselves as they separate.  This flux
tube, which describes the highly excited QCD vacuum, decays via a
nonperturbative particle-antiparticle production process analogous to the
Schwinger mechanism.

Particle production by the flux tube is described by a source term and in
quantum field theory that source term is
non-Markovian~\cite{kme,basti,gsi,prd,bloch,memory}; i.e., essentially
nonlocal in time.  This feature can be important when the fields are strong.
(In the weak-field limit a time-local Schwinger-like source term is
recovered.)  Such a situation is plausible at RHIC and especially at
LHC~\cite{bastirev,memory} where the anticipated initial energy densities
are, respectively, $\varepsilon \sim 10$--$100\,$GeV/fm$^3$ and $\varepsilon
\gtrsim 1\,$TeV/fm$^3$.

Another feature that is characteristic of the flux-tube production mechanism
is the back-reaction phenomenon.  This phrase simply describes the fact that
once the particles are produced they are accelerated and thereby generate a
field that interferes with the collisional field that produced them.  Plasma
oscillations are then almost inevitable~\cite{Back}, although they can
rapidly be damped if the thermalising collisions between particles are
frequent~\cite{bloch,memory,KM,bahl,bahl2,eis,rta,baym,nayak2}.

An observable that may preserve information about non-Markovian effects and
plasma oscillations in the pre-equilibrium particle-production stage of an
URHIC is the thermal dilepton spectrum because leptons do not participate in
the strong interactions that equilibrate the QGP~\cite{lerran}.  To explore
that possibility, we employ a quantum Vlasov equation with a non-Markovian
source term to calculate the single particle distribution function that
characterises particle production by the flux tube, $f(\vec{p},t)$, and use
that to calculate the dilepton spectrum and its evolution from impact to
equilibration.

Of course, equilibration can only be effected by dissipative processes, such
as collisions, and herein we describe those effects via a relaxation time
approximation (RTA).  This is a coarse representation of the interactions
between the partons produced in the collision but, even so, ensuring
thermodynamic consistency is nontrivial.  We introduce and describe one
practical scheme for achieving that goal.

Our article is organised as follows.  In Sec.~\ref{sec:II} we present the
quantum kinetic equation and review properties of the source term.  The
collision term is described in Sec.~\ref{sec:III}.  This completes the
specification of the model and so our results appear in Sec.~\ref{sec:IV}.
Section~\ref{sec:V} presents some concluding remarks.

\section{Distribution Function}
\label{sec:II}
\subsection{Kinetic Equation and Source Term}
\label{subsec:IIaa}
We assume that in its wake an URHIC leaves a high energy-density electric
field, which occupies a large (in fact, unbounded) spacetime volume.  This
excited domain decays via a Schwinger-like mechanism, producing an
unequilibrated plasma of highly energetic quarks and antiquarks.  That system
evolves and equilibrates, forming a component of the QGP.  Hadronisation only
takes place at a later stage, when the temperature and density of the
equilibrated system fall below some critical values, and herein we do not
consider that process.

We represent the excited domain by a spatially homogeneous, time-dependent
Abelian vector potential, $A_\mu(t)$, and work in the temporal gauge:
$A_0=0$.  The spatial part of the vector potential defines the $\hat
z$-direction; i.e., $\vec{A}(t) = (0,0,A(t))$, and generates an electric
field $\vec{E}(t) = -d\vec{A}(t)/dt$.  This provides the input for the
quantum kinetic equations whose solution describes the evolution of the
single parton distribution functions that characterise the produced partons:
\begin{eqnarray}
\label{KE}
\frac{d}{dt} f_\pm(\vp,t) & = & S_\pm(\vp,t)+C_\pm(\vp,t) \,\,,\\
\mbox{where}\;\;\frac{d}{dt} & := &
\frac{\partial }{\partial t}+eE(t)\frac{\partial }{\partial p_\parallel}\,,
\end{eqnarray}
$e$ is the electric charge and $+/-$ denotes bosons/fermions; and
$f_\pm(\vp,t)$ gives the ensemble fraction of particles with a given
momentum, $\vp$, at time $t$.  The analogous equation for antiparticles,
which is obtained via charge conjugation and must be solved simultaneously,
yields $\bar f_\pm(\vp,t)$, the single antiparticle distribution function.
The feedback generated by the motion of the partons is incorporated by
coupling Maxwell's equation to Eq.~(\ref{KE}) and its analogue, as we discuss
in Sec.~\ref{subsec:IIa}.  That also introduces a coupling between the
equations for $f$, $\bar f$.

In Eq.~(\ref{KE}), $C_\pm(\vp,t)$ is the collision term, which also couples
the equations for $f$ and $\bar f$, and which we discuss in
Sec.~\ref{sec:III}; and $S_\pm(\vp,t)$ is the particle-antiparticle-producing
source term:
\begin{eqnarray}
S_\pm(\vp,t)& = & \frac{1}{2}W_\pm(\vp,t,t)\!\! \int_{t_0}^t\! dt'\,
W_\pm(\vp ,t,t')\big[1\pm 2f_\pm(\vp,t')\big]\nonumber \\
&&\times\cos\left[ 2\int_{t'}^t d\tau ~\omega (\vp,t,\tau) \right].
\label{source} 
\end{eqnarray}
The effect of quantum statistics on the particle production rate is evident
in the ``$ \pm 2 \,f_\pm$'' in Eq.~(\ref{source}) (neglecting this term
defines the low density limit) and in the different transition (or
tunnelling) amplitudes
\begin{equation}
\label{W}
{\cal W}_\pm(\vp,t,t')=eE(t')\frac{p(t,t')}{\omega^2 (\vp,t,t')}
\left(\frac{\varepsilon_\perp}{p(t,t')} \right)^{g_\pm - 1}\,,
\end{equation}
where: 
$g_\pm= 2 s_\pm + 1$, with $s_+=0$, $s_-=1/2$;
the three-vector momentum $\vp=(\vp_{\perp} ,p_\parallel)$; the transverse
mass-squared $\varepsilon_{\perp}^2=m^2+p_{\perp}^2$;
$\omega^2(\vp,t,t')=\varepsilon_{\perp}^2+p^2(t,t')$; and
\begin{equation}
\label{kinmom}
p(t,t')=p_\parallel - e\,[A(t)-A(t')]
=p_\parallel+e \int_{t'}^{t} d\tau~E(\tau)\,.
\end{equation}
(Equation~(\ref{kinmom}), which describes the action of the field on the
particles, is just a re-expression of the Lorentz force law: $ \partial
p(t,t^\prime)/\partial t = e E(t)$.)  The source term is nonlocal in time and
that can be important on short time-scales in strong fields: if the fields
are strong enough the time duration of a tunnelling event and the time
between successive events, which is set by the particles' Compton
wavelengths, are similar, and the processes interfere, with observable
consequences in the distribution function.  In addition, strong fields
enhance the differences between fermion and boson production.  These features
were highlighted in Refs.~\cite{kme,prd}.

With our initial setup we have a simple, algebraic symmetry between the
particle and antiparticle distribution functions:
\begin{equation}
\bar f(\vec p,t) = f(-\vec p,t)\,,
\end{equation}
and hence it is only necessary to consider Eq.~(\ref{KE}) explicitly.  To
simplify this kinetic equation we follow Ref.~\cite{bloch} and introduce two
real auxiliary functions
\begin{eqnarray}
u_\pm(\vp,t)&=&\int_{t_0}^t\! dt'\, W_\pm(\vp ,t,t') \big[1\pm
2f_\pm(\vp,t')\big]\nonumber \\
&&\times \sin\left[ 2\int_{t'}^t d\tau \,\omega (\vp,t,\tau) \right],\\
v_\pm(\vp,t)&=&\int_{t_0}^t dt' W_\pm(\vp ,t,t')\big[1\pm
2f_\pm(\vp,t')\big]\nonumber \\ 
&&\times \cos\left[ 2\int_{t^\prime}^t d\tau \,\omega (\vp,t,\tau)
\right],
\end{eqnarray}
such that for $C_\pm(\vp,t)=0$: 
$ u_\pm^2+ v_\pm^2 \mp \, (1\pm 2f)^2= {\rm const.}$,
with the initial conditions $f(t_0)=v(t_0)=u(t_0)=0 $.  This permits us to
rewrite Eq.~(\ref{KE}) as a system of coupled, first-order differential
equations
\begin{eqnarray}
\label{KEdiff}
\frac{d}{dt} f_\pm &=& \frac12 { W_\pm(\vp,t)}\,v_\pm\,+\,C_\pm(\vp,t) \,,\\
\frac{d}{dt} u_\pm &=& 2 \, \omega(\vp)\, v_\pm\ ,\\
\frac{d}{dt} v_\pm &=& \ds \, W_\pm(\vp,t) \,[1\pm 2 f_\pm] -
2\,\omega(\vp)\, u_\pm \,,
\label{KEdiff3}
\end{eqnarray}
where $W_\pm(\vp,t)$ and $\omega(\vec{p})$ denote, respectively,
$W_\pm(\vp,t,t)$ and $\omega(\vec{p},t,t)$.  While this complex is
qualitatively identical to Eq.~(\ref{KE}), it is simpler to treat
numerically.

The kinetic equation describes pair creation for both bosons and fermions.
In the following we identify the fermionic degrees of freedom with the quarks
and antiquarks produced in an URHIC and restrict ourselves solely to the
fermionic case, henceforth suppressing the $\pm$ subscript.

\begin{figure}[t]
\centerline{\includegraphics[height=6.5cm]{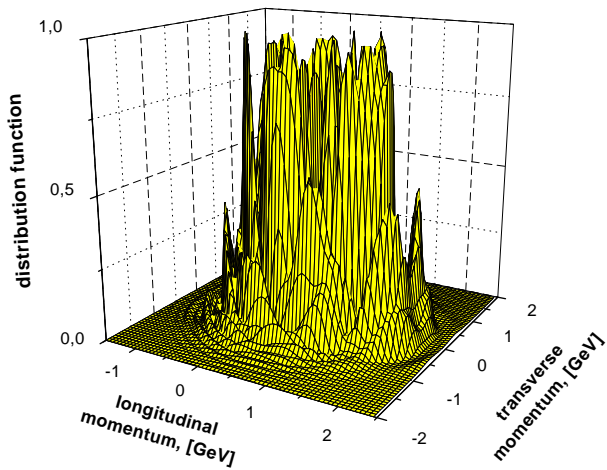}}
\caption{\label{3d1} Momentum dependence of the single parton distribution
function: $f(\vp,t=10\,{\rm fm})$, obtained by solving
Eqs.~(\protect\ref{KEdiff})-(\protect\ref{KEdiff3}) and
(\protect\ref{1.14})-(\protect\ref{Etot}), with the impulse profile in
Eq.~(\protect\ref{Eex}) ($A_0= 1\,$GeV$^2$, $1/b= 1\,$GeV, and $m b = 1/5$;
i.e., Set~4 in Table~\ref{tableA}) and neglecting collisions.  ``transverse
momentum'' represents $|\vec{p}_\perp|$ and ``longitudinal momentum''
represents $p_\|$.  The irregular structure makes clear that this is not the
distribution function of a system in equilibrium.}
\end{figure}

\begin{figure}[t]
\centerline{\includegraphics[height=6.7cm]{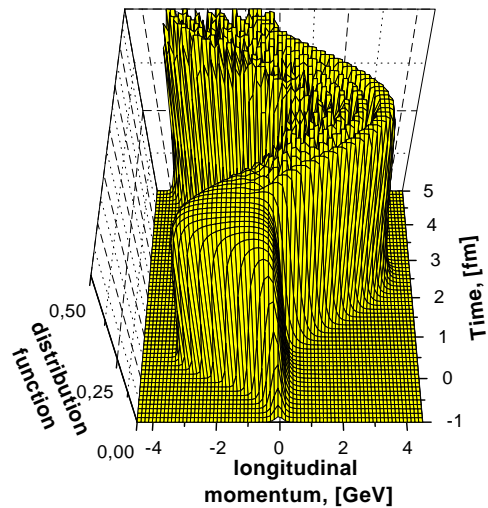}}
\caption{\label{3d2} Time evolution of the single parton distribution
function: $f(p_\perp=0,p_\|,t)$ in the low density limit.  (Model parameters
are as in Fig.~\protect\ref{3d1}.)  The regular (plasma) oscillation is
driven by the back-reaction phenomenon.}
\end{figure}

\subsection{Maxwell's Equation and Internal Currents}
\label{subsec:IIa}
The kinetic equation depends nonlinearly on the time-dependent electric
field.  The URHIC provides the impetus for this field: it provides an
external field, which we model via
\begin{equation}
\label{Eex}
E_{ex}(t) = -A_0\, {\rm sech}^{2}(t/b)\,.
\end{equation}
This profile ``switches-on'' at $t\sim -2 b$ and off at $t\sim 2 b$, and
attains its maximum magnitude of $A_0$ at $t=0$.  The external field,
$E_{ex}$, polarises the vacuum, generating a polarisation current that
depends on the dielectric properties of the medium, which are encoded in the
source term, and promotes the spontaneous production of particle-antiparticle
pairs, which it then accelerates, generating a conduction current that
depends on the particle distribution function.  A consequence of the URHIC
then is the appearance of a two component internal current and an attendant
internal electric field:
\begin{equation}
\label{1.14}
- {\dot E}_{in}(t) = j_{in} = j_{cond}(t) + j_{pol}(t)\,,
\end{equation}
where the fully renormalised currents are~\cite{bloch}
\begin{eqnarray}
\label{jcond}
j_{cond}(t) & = & 2\, N_p\,e \!
\int\!\frac{d^3p}{(2\pi)^3}\,\frac{p_\|}{\omega(\vp)}\, f(\vp,t) \,,\\
%
%
%
\nonumber j_{pol}(t) & = & N_p\,e\, \!\int\!\frac{d^3p}{(2\pi)^3}\, 
\frac{\varepsilon_\perp}{\omega(\vp)} 
\left[ v(\vec p,t) - \,\frac{e \dot E(t)\,\varepsilon_\perp }{4\,
\omega^4(\vp)} \right],\\
\label{jpol}
\end{eqnarray}
with $N_p=18$, as described in Table~\ref{tableA}.

\begin{figure}[t]
\centerline{\includegraphics[height=6.7cm]{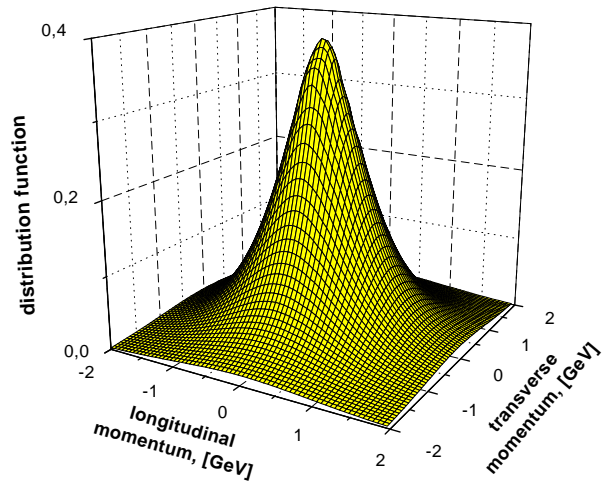}}
\caption{\label{3d3} $f(\vec{p},t=10\,{\rm fm})$ obtained with the inclusion
of parton-parton collisions, as described in Sec.~\protect\ref{sec:III}.
Collisions promote equilibration of the system, which is evident in the now
smooth distribution function cf.\ Fig.~\protect\ref{3d1}.  (Model
parameters are as in Fig.~\protect\ref{3d1}.)}
\end{figure}

In the absence of collisions, Eqs.~(\ref{KEdiff})-(\ref{KEdiff3}) and
(\ref{1.14})-(\ref{jpol}), with
\begin{equation}
\label{Etot}
E(t) = E_{ex}(t) + E_{in}(t)\,,
\end{equation}
form a closed system of coupled equations whose solution provides the
time-dependent electric field and single parton distribution function.  We
can use this system to illustrate some of the effects we have mentioned.  In
Fig.~\ref{3d1} we depict the momentum dependence of the distribution
function.  The irregular structure is produced by interference effects in the
non-Markovian source term, which arise because the tunnelling time is of the
same magnitude as the Compton wavelength of the produced particles, and by
the feedback mechanism.  This structure is averaged out in the ideal Markov
limit~\cite{kme} and makes very clear that this is {\bf not} the distribution
function of a system in equilibrium.  Figure~\ref{3d2} depicts the time
evolution of the calculated distribution function and the feedback that
characterises the behaviour of the internal currents is manifest in the
obvious, regular plasma oscillation.  For comparison, in Fig.~\ref{3d3} we
depict the distribution function obtained after the inclusion of dissipative
effects, to be discussed in Sec.~\ref{sec:III}.

\begin{table}[t]
\caption{\label{tableA} Parameter sets used to specify our model of an URHIC,
Eq.~(\protect\ref{Eex}).  They yield equilibrium values of thermodynamic
quantities (three rightmost columns) that are consistent with those expected
in a QGP, with Sets~3, 4 approximating RHIC-like conditions.  We use $m b =
1/5$, a strong coupling: $e=1$, $\tau_c = 1$, and consider $3$ quark
flavours, which explains the factor of $N_p = 18 = 2_{\rm spin} \, 3_{\rm
flavour}\,3_{\rm colour}$ that appears frequently.}\vspace*{1ex}
\begin{ruledtabular}
\begin{tabular*}
{\hsize}
{l|@{\extracolsep{0ptplus1fil}}c@{\extracolsep{0ptplus1fil}}c|@{\extracolsep{0ptplus1fil}}c@{\extracolsep{0ptplus1fil}}c@{\extracolsep{0ptplus1fil}}c@{\extracolsep{0ptplus1fil}}c}
    &$A_0$ [GeV$^2$]&$b$ [fm]&$\varepsilon$ [GeV/fm$^{3}$]&$T$ [GeV]&$n$ [fm$^{-3}$]\\\hline
Set 1& 0.25         &0.2     &1.0               &0.20     &1.4\\
Set 2& 0.40         &0.2     &2.5               &0.26     &3.0\\
Set 3& 0.75         &0.2     &12~~              &0.38     &10~~ \\
Set 4& 1.0~         &0.2     &24~~              &0.44     &17~~
\end{tabular*}
\end{ruledtabular}
\end{table}

\section{Equilibrating collisions}
\label{sec:III}
Once the particles are produced they are accelerated by the electric field
and, as we saw in Sec.~\ref{subsec:IIa}, if $e E \sim \varepsilon_\perp^2$
then large amplitude, high frequency plasma oscillations appear.  This
collective effect, which is a hallmark of our approach, may have observable
consequences in experiments aimed at producing an equilibrated QGP.  Whether
that is the case or not can only be determined once the effect of
parton-parton collisions is incorporated.

The general nature of the dissipative collision term is known: it too is
non-Markovian and can produce particles~\cite{CGreiner}.  However, its
complexity mitigates against its use in semi-quantitative, exploratory
studies and hence herein we employ a simple
RTA~\cite{bloch,memory,bahl,bahl2,eis,rta,baym,nayak2}.  

In this approach the detailed description of parton-parton scattering is
replaced by a continuous viscosity term, which involves a time-dependent
parameter that is identified with the collision period.  In addition we
suppose that the thermodynamic laws are valid at each time $t$, and this
assumption of local-equilibrium provides for an internally consistent
definition and calculation of time-dependent thermodynamic variables, such as
temperature and energy density.  Of course, accepting a physical
interpretation of these quantities only makes sense once the violent effects
of the URHIC have subsided and the quantities are evolving slowly with $t$.

The collision term in the kinetic equation for $f(\vec p,t)$ must describe
particle-particle $(pp)$ and particle-antiparticle $(pa)$ collisions and
reflect the symmetries of our initial conditions.  Hence we employ
\begin{eqnarray}
\nonumber \lefteqn{C(\vp,t;T,u^{\nu}) = }\\
&& \nonumber \frac{1}{\tau_{pp}(t)} \left[
f_{eq}\big(\vp,t;T(t),u^{\nu}(t)\big) - f \big(\vp,t\big)\right]\\
&& + \frac{1}{\tau_{pa}(t)} \left[ f_{eq}\big(- \vp,t;T(t),u^{\nu}(t)\big)
- f \big(\vp,t\big)\right]
\label{coll}
\end{eqnarray}
where $\tau_{pp}(t)= \tau_{pa}(t) = \tau(t)$ is the time-dependent relaxation
time and
\begin{equation}
\label{eq}
f_{eq}\left(\vp,T,u^{\nu}\right)= \left[ \exp\left(
\frac{p_{\nu}u^{\nu}(t)}{T(t)}\right) + 1 \right]^{-1} 
\end{equation}
is the quasi-equilibrium distribution function, with $p_{\nu}$ the quarks'
4-momentum.  The other quantities in Eq.~(\ref{coll}) are:
\begin{equation}
\label{ut}
u^{\nu}(t)= (1,0,0,u(t))\,[1-u(t)^2]^{-(1/2)}\,,
\end{equation}
the hydrodynamical four-velocity; and $T(t)$, the local-equilibrium
temperature, both of which we define below.

Our RTA encodes all the complicated effects of parton-parton collisions in a
single time-dependent quantity, the relaxation time: $\tau(t)$.  It is a
measure of the time between successive collisions and as such can be
expressed as~\cite{nayak2}
\begin{equation}
\label{relax}
\tau(t) = \tau_c\, \frac{\lambda(t)}{\bar\upsilon(t)}\,,
\end{equation}
where $\lambda(t)=1/[n(t)]^{1/3}$ is the mean interparticle separation, since
\begin{equation}
\label{ndensity}
n(t) = N_p\, \int\!\frac{d^3 p}{(2\pi)^3} \, f(\vp,t)
\end{equation}
is the mean particle number density, and 
\begin{equation}
\bar\upsilon(t)= |\vec p|_f(t)/\varepsilon_f(t)
\end{equation}
is an average speed.  (See Eqs.~(\ref{edensity}) and (\ref{avemagp}).) The only
parameter in our implementation of the relaxation time approximation is then
$\tau_c$, the dimensionless constant of proportionality.

Returning to Eq.~(\ref{coll}), we define the local-temperature by requiring
that at each time, $t$, the mean particle energy density in the plasma is
identical to that in an equilibrated plasma at a temperature $T(t)$; i.e.,
\begin{eqnarray}
\label{edensity} \varepsilon_f(t) & = & N_p
\,\int\!\frac{d^3p}{(2\pi)^3}\,\omega(\vec{p})\, 
\left[f(\vp,t)-z_f(\vp,t)\right]\\
& = & N_p \, \int\!\frac{d^3p}{(2\pi)^3}\,\omega(\vec{p})\, f_{eq}(\vp,t) \,.
\label{econs}
\end{eqnarray}
Similarly, $u(t)$ in Eq.~(\ref{ut}) is defined via the requirement
\begin{eqnarray}
\label{momdensity} \vec p_f(t)&= & N_p\,\int\!\frac{d^3p}{(2\pi)^3}\, \vp \,
\left[ f(\vp,t)-z_f(\vp,t)\right]\\
& = & N_p\,\int\!\frac{d^3p}{(2\pi)^3}\,\vp\, f_{eq}(\vp,t) \,;
\label{pcons}
\end{eqnarray}
i.e., that the mean particle three-momentum is the same as that of an
equilibrated plasma characterised by an hydrodynamical particle velocity
$u(t)$.  The average magnitude of the momentum is
\begin{equation}
\label{avemagp}
|\vec p|_f(t) = N_p\,\int\!\frac{d^3p}{(2\pi)^3}\, |\vp| \,
\left[ f(\vp,t)-z_f(\vp,t)\right]\,.
\end{equation}
The new element in these equations,
\begin{eqnarray}
\nonumber z_f(\vp,t) & = &\bigg(\frac{e\,\varepsilon_{\perp}}{4
\,\omega^3(\vec p)}\bigg)^2 \bigg[ E^2(t) \\
&& - \frac{2}{\tau_c}\,e^{-2t/\tau_c}\,\int_{t_0}^{t}dt' E^2(t')
e^{2t'/\tau_c} \bigg]\,,
\label{counter}
\end{eqnarray}
is a regularising counterterm, determined via the same procedure~\cite{bloch}
that yields the renormalised currents in Sec.~\ref{subsec:IIa}, which ensures
that the integrals involving the calculated distribution function are finite.
This counterterm itself exhibits ``memory effects;'' i.e., it is sensitive to
the time-history of the electric field.

We judge that using a time-dependent relaxation time is an improvement over
previous work that used $\tau(t)=\,$const.; e.g., Refs.~\cite{bloch,memory},
because it is manageable and better models the conditions produced by an
URHIC.  Clearly, just after the impact the parton number density is small and
hence the time between successive collisions is large.  Over time, however,
the density of produced partons increases, leading to a reduction in the
interval between collisions.  These features are crudely reflected by the
evolution of the relaxation time described in Eq.~(\ref{relax}).  

\begin{figure}[t]
\centerline{\includegraphics[height=5.0cm]{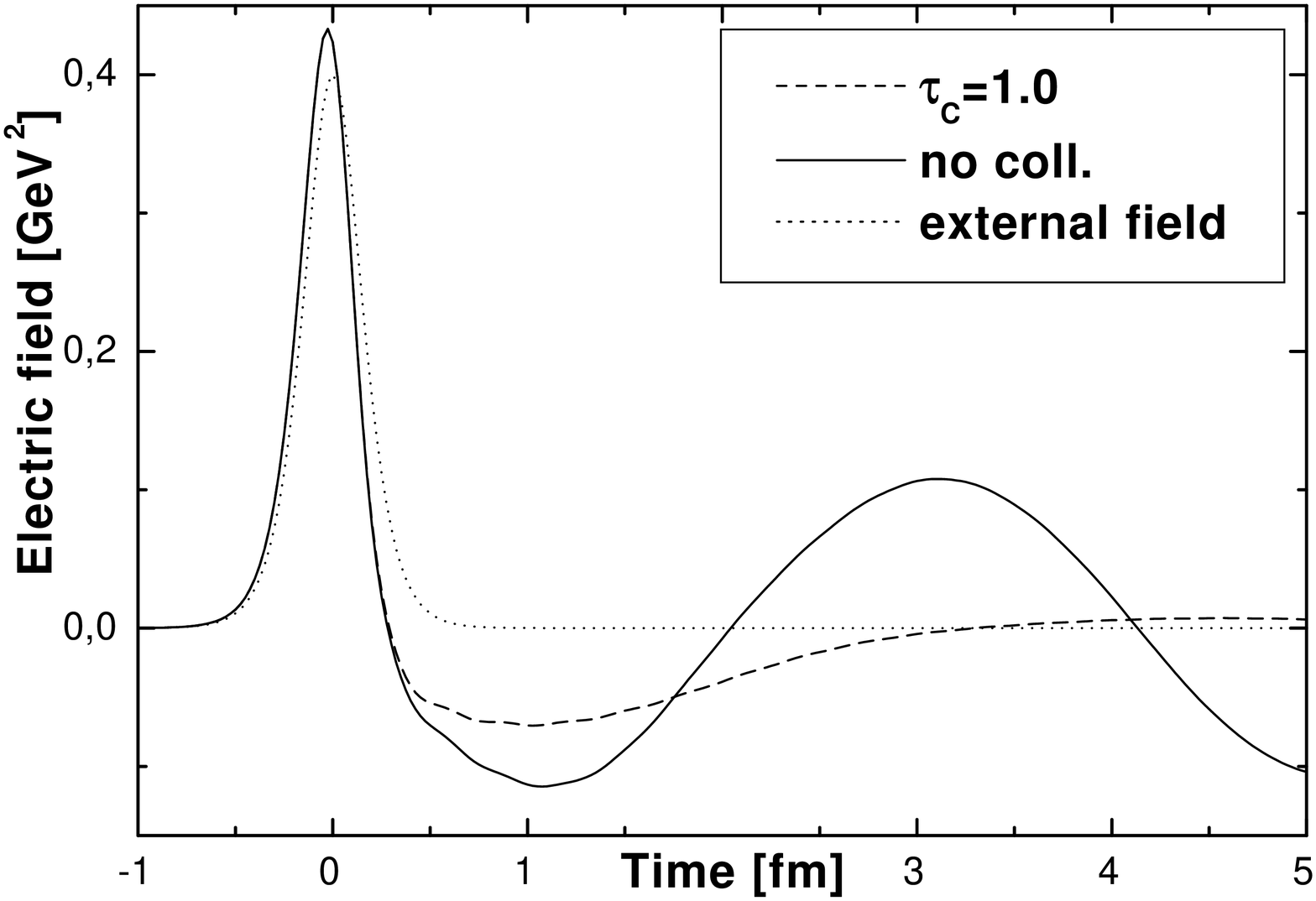}}
\caption{\label{elec} Time evolution of the total electric field.  The
effects of feedback are evident in the oscillatory behaviour of $E(t)$.  The
viscous collision term damps these oscillations in a characteristic time
$\tau_c/m \sim 1\,$fm.  The external impulse electric field is also depicted.
(Parameters: Set~2, Table~\protect\ref{tableA}.)}
\end{figure}

\begin{figure}[ht]
\centerline{\includegraphics[height=5.0cm]{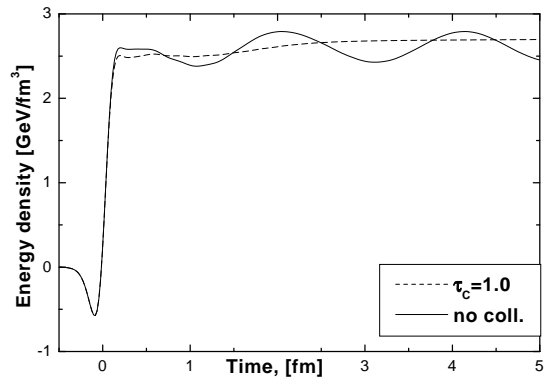}}
\caption{\label{energ} Time evolution of the calculated energy density,
Eq.~(\protect\ref{edensity}).  It is negative in the neighbourhood of $t=0$
because of the large vacuum polarisation produced by the strong external
field (the URHIC).  This feature is quickly compensated by particle
production in the aftermath of the impact.  Collisions subsequently yield an
equilibrated QGP.  (Parameters: Set~2, Table~\protect\ref{tableA}.)}
\end{figure}

The final form of our collision term is thus obtained from the combination of
Eqs.~(\ref{coll}) and (\ref{relax}); i.e.,
\begin{eqnarray}
C(\vec{p},t) & = & \frac{\bar\upsilon(t)}{\lambda(t)} \,\left(
\frac{f_{eq}(\vec{p},T(t),u(t))- f(\vec{p},t)}{\tau_c} \right.\nonumber \\
&& + \left. \frac{f_{eq}(-\vec{p},T(t),u(t))- f(\vec{p},t)}{\tau_c}
\right)\,.
\label{collF}
\end{eqnarray}
It provides additional nonlocal feedback in the solution for $f(\vec{p},t)$.
We emphasise again that our RTA is based on the assumption of
local-equilibrium, which is valid for $|t|\gg b$, where $b$ is the time
duration of the URHIC.  For $|t|\lesssim b$, however, it is of questionable
validity and may lead to model-dependent artefacts, the misinterpretation of
which one must guard against.

\begin{figure}[t]
\centerline{\includegraphics[height=5.0cm]{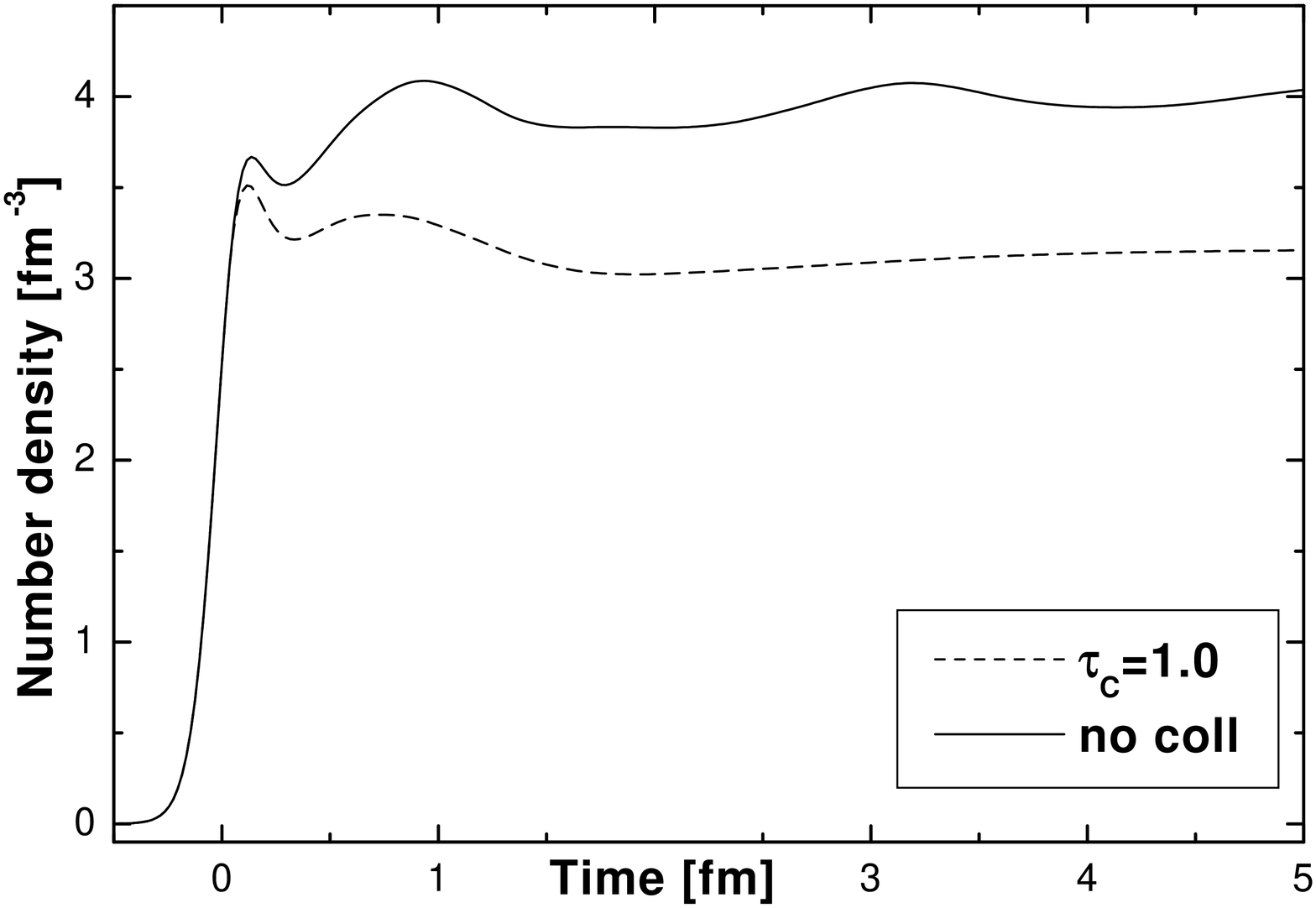}}
\caption{\label{number} Time evolution of the particle number density,
Eq.~(\protect\ref{ndensity}).  The equilibrium value: $3.0\,$fm$^{-3}$, is
that listed in Table~\ref{tableA}.  (Parameters: Set~2,
Table~\protect\ref{tableA}.)}
\end{figure}

\begin{figure}[th]
\centerline{\includegraphics[height=5.0cm]{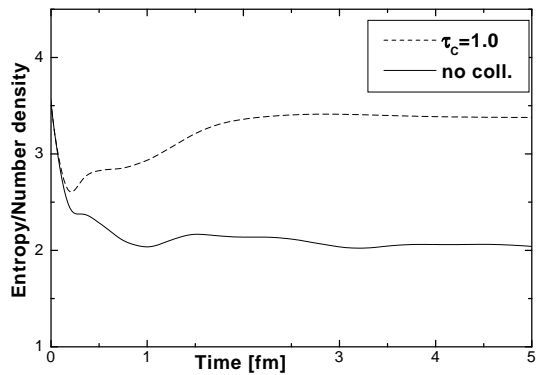}}
\caption{\label{entro} Time evolution of the entropy/particle,
Eq.~(\protect\ref{entropy}), with and without collisions.  It is intuitively
obvious why the entropy is larger when particle-particle collisions are
incorporated: in equilibrating the system, collisions increase the degree of
disorder.  (Parameters: Set~2, Table~\protect\ref{tableA}.)}
\end{figure}

\section{Numerical results}
\label{sec:IV}
\subsection{Thermodynamic Parameters}
\label{subsec:IVa}
Our results are obtained via the simultaneous solution of
Eqs.~(\ref{KEdiff})-(\ref{KEdiff3}), (\ref{1.14})-(\ref{jpol}),
(\ref{econs})-(\ref{counter}), using the collision term in Eq.~(\ref{collF}),
which we accomplish using a fourth-order Runge-Kutta procedure.  The
solution, $f(\vec{p},t)$, fully describes the plasma's evolution, from its
creation to equilibration.  In addition we obtain the time-dependent vector
potential, electric field and currents, and also the quasi-equilibrium
temperature and collective velocity.

\begin{figure}[t]
\centerline{\includegraphics[height=5.0cm]{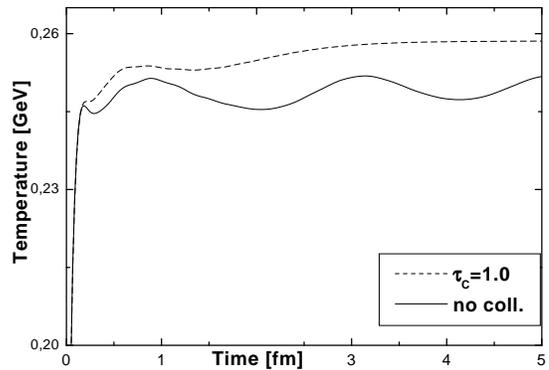}}
\caption{\label{temper} Time evolution of the quasi-equilibrium temperature.
The equilibrium value reached here: $0.26\,$GeV, is greater than the
anticipated critical temperature for QGP formation.  (Parameters: Set~2,
Table~\protect\ref{tableA}.)}
\end{figure}

\begin{figure}[ht]
\centerline{\includegraphics[height=5.0cm]{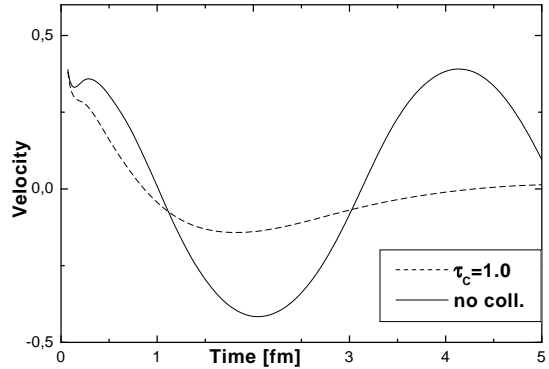}}
\caption{\label{veloc} Time evolution of the collective velocity, $u(t)$.
It's behaviour clearly signals the collective plasma oscillation that is
characteristic of the flux-tube production mechanism.  The plasma oscillation
is damped by collisions.  (Parameters: Set~2, Table~\protect\ref{tableA}.)}
\end{figure}

To explore the solution's properties we have employed a range of parameter
sets, which are listed in Table~\ref{tableA}, and in
Figs.~\ref{elec}--\ref{veloc} we demonstrate the effect of collisions by
comparing the solution obtained using $\tau_c=1.0$ with that obtained in the
collisionless limit $\tau_c\to \infty$.  (NB.\ As evident in the figures, the
plasma period satisfies $m \,\tau_{pl} \approx 2.5 \sim \tau_c$.  For
$\tau_c\ll m\,\tau_{pl}$ plasma oscillations are not observable~\cite{bloch};
i.e., the system is overdamped.)

In Fig.~\ref{elec} we see that the URHIC generates a strong electric field,
which produces particles that sustain the field for a time that depends on
the collision frequency: a large value of $\nu_c = 1/\tau_c$ means a
short-lived electric field and rapid equilibration.

Figure~\ref{energ} depicts the time evolution of the energy density,
Eq.~(\ref{edensity}).  It is negative on a small domain around $t=0$ because
of the large vacuum polarisation induced by the strong external field.  This
instability is quickly corrected by rapid particle creation; a correlation
that is apparent in a comparison of Fig.~\ref{energ} with Fig.~\ref{number},
which portrays the particle number density, Eq.~(\ref{ndensity}).  The
density reaches a higher value in the absence of collisions because the
field-current feedback allows unhindered, repeated bursts of particle-pair
creation.  The Pauli principle does not significantly retard the process
because, while the particles are preferentially produced with small momenta,
the field rapidly accelerates them.  As Table~\ref{tableA} shows, the number
density increases with increasing $A_0$; i.e., with increasing impulse field
strength.

In Fig.~\ref{entro} we plot the behaviour of the entropy/particle, where
the entropy density is:
\begin{equation}
\label{entropy}
s(t) = - N_p\,\int\! \frac{d^3p}{(2\pi)^3}\, f(\vp,t)\ln f(\vp,t)\,.
\end{equation}

The calculated quasi-equilibrium temperature is depicted in
Fig.~\ref{temper}.  After rising quickly, it settles into a slow evolution
once the external field, Eq.~(\ref{Eex}), has subsided.  This marks the
beginning of the domain on which the concept of local equilibrium is valid.

The local velocity is plotted in Fig.~\ref{veloc}.  It shows, as one would
intuitively expect, that the produced particles are accelerated, reaching
their maximum velocity when the electric field vanishes (cf.\
Fig.~\ref{elec}), then decelerated as the field reverses direction.  They
stop, and then reverse direction and are accelerated by the reversed field to
a new maximum velocity.  The repetition of this pattern is the collective
plasma oscillation, which is clearest in this figure.  Of course, in
equilibrating the system, collisions act to destroy the pattern.

\begin{figure}[t]
\centerline{\includegraphics[height=5.5cm]{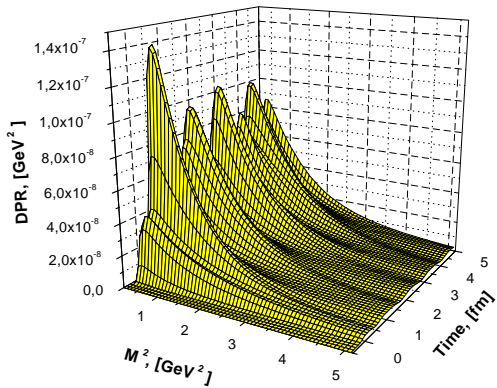}}
\caption{\label{dilep0} Time evolution of the dilepton production rate.  The
plasma oscillation is evident in this collisionless case, as is the feature
that the dileptons are preferentially produced at low-$M^2$.  (Parameters:
Set~4, Table~\protect\ref{tableA}.)}
\end{figure}

\subsection{Dilepton Production}
\label{subsec:IVb}
The plasma oscillation is evident in the electric field and in each of the
thermodynamic variables but none of these quantities are directly measurable.
Is there any way that this characteristic signature of the flux-tube
production mechanism can be observed?  

The thermal dilepton production rate may provide a means.  Dileptons produced
in the URHIC do not interact strongly and hence those produced soon after the
impact carry and transmit information about the pre-equilibrium stage of the
plasma.  The dilepton production rate from our (quasi-)equilibrium three
quark system can be estimated using~\cite{lerran,wongHIC}
%
%
%
\begin{eqnarray}
&& \frac{dN}{dt\,d^3x\,dM^2} = \nonumber \\
&& \frac{\alpha^2}{3\pi^3}\,(1- 4 m_l^2/M^2)^{\frac{1}{2}} \,
\bigg(1+2\frac{m^2+m_l^2}{M^2}+4\frac{m^2m_l^2}{M^4}\bigg) \, \nonumber \\ 
&& \times \int_{m}^\infty \!\! d\epsilon_1 d\epsilon_2 \, f(\epsilon_1)\,
f(\epsilon_2)\, \theta(M^2-M^2_-) \, \theta(M^2_+-M^2)\,,\nonumber \\
\label{dile}
\end{eqnarray}
where: $\alpha = 1/137$; $m_l$ is the lepton mass, $m$ the quark mass and we
use $m_l=m$; $M^2$ is the invariant mass of the produced dilepton pair; 
\begin{equation}
M^2_\pm = 2\, m^2 + 2 \,\epsilon_1 \epsilon_2 \pm 2\, |\vp_1|\, |\vp_2| \,;
\end{equation}
and the time-dependence of the distribution functions is implicit.

\begin{figure}[t]
\centerline{\includegraphics[height=5.5cm]{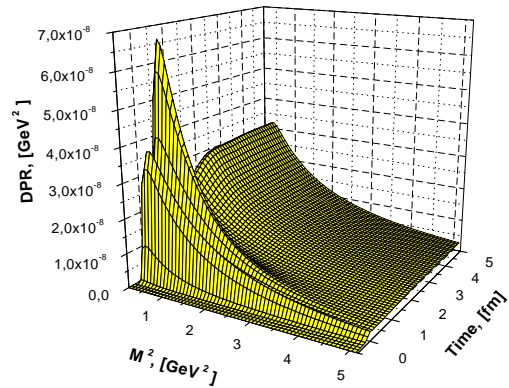}}
\caption{\label{dilep1} Time evolution of the dilepton production rate.
Frequent collisions rapidly damp the plasma oscillation, forcing the
production rate to settle at a constant value. (Parameters: Set~4,
Table~\protect\ref{tableA}.)}
\end{figure}

We obtain the distribution functions, $f(\epsilon_{1,2})$ in
Eq.~(\ref{dile}), as described in Sec.~\ref{subsec:IVa}, using the RHIC-like
parameter-Set-4 in Table~\ref{tableA}.  Knowing them, the calculation of the
dilepton production rate is straightforward.  In Fig.~\ref{dilep0} we display
that rate as a function of $(M^2,t)$.  The plasma oscillation is evident in
the time evolution, with more dilepton pairs being produced when the electric
field is strongest and the parton production rate peaks, and no pairs being
produced when the electric field vanishes.  The effect of collisions is to
drive the system to equilibrium where the thermal dilepton production rate
becomes constant, as shown in Fig.~\ref{dilep1}.

Clearly, the plasma oscillations generate a signal.  However, the time
evolution of the dilepton production rate is a difficult quantity to measure.
An easier quantity is the time-integrated rate:
\begin{equation}
\label{tint}
\rho_{l^+l^-}(t) := \frac{dN}{d^3x\,dM^2} =
\int_0^t\!dt'\,\frac{dN}{dt'\,d^3x\,dM^2}\,.
\end{equation}
Does a signal survive in this observable?

\begin{figure}[t]
\centerline{\includegraphics[height=5.5cm]{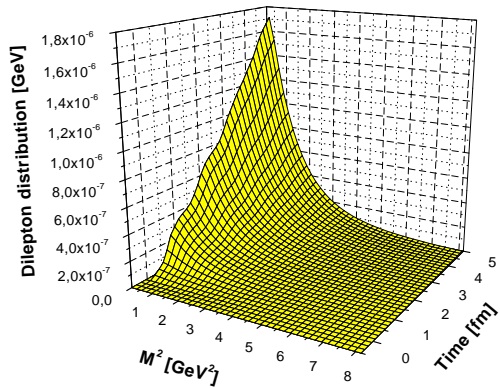}}
\caption{\label{dilep3} Time integrated dilepton production rate,
Eq.~(\protect\ref{tint}), calculated without collisions and hence in the
presence of a persistent plasma oscillation.  (Parameters: Set~4,
Table~\protect\ref{tableA}.)}
\end{figure}

\begin{figure}[t]
\centerline{\includegraphics[height=5.5cm]{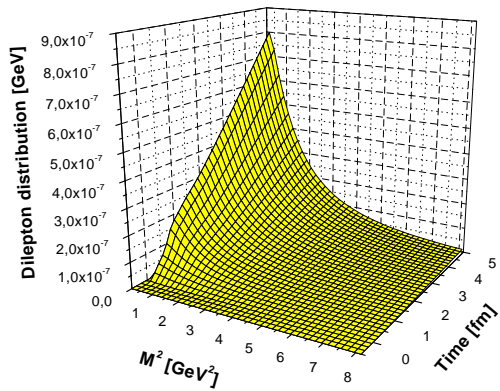}}
\caption{\label{dilep4} Time integrated dilepton production rate,
Eq.~(\protect\ref{tint}), calculated under the influence of frequent
collisions that rapidly equilibrate the plasma.  (Parameters: Set~4,
Table~\protect\ref{tableA}.)}
\end{figure}

In Fig.~\ref{dilep3} we plot $\rho_{l^+l^-}(t)$ obtained in the absence of
collisions, which must be compared with the function in Fig.~\ref{dilep4}
that was obtained with the inclusion of collisions.  Our simple model yields
rates that are comparable with other estimates; e.g., Ref.~\cite{dutta}, and
the comparison of the figures shows that plasma oscillations generate an
enhancement in the number of dileptons, $\rho_{l^+l^-}(t)$, which is as large
as a factor of $2$ at $t=5\,$fm, given the Set~4 initial conditions in
Table~\ref{tableA}.  (Note that collisions eliminate the plasma oscillation
so Fig.~\ref{dilep4} can be thought of as the oscillation-free scenario.)

\begin{figure}[ht]
\centerline{\includegraphics[height=5.7cm]{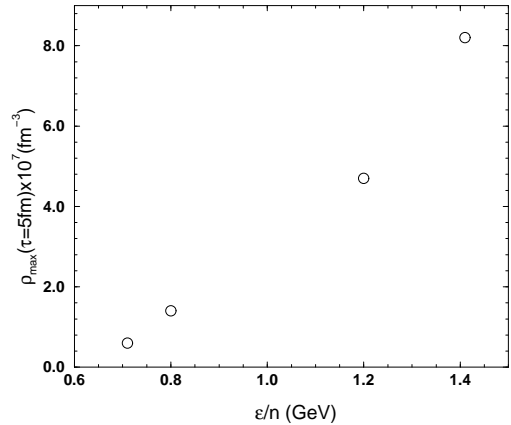}}
\caption{\label{Ntaufig} $\rho_{l^+l^-}(t=5\,{\rm fm})$ at the low invariant
mass, $M^2$, for which it takes its maximum value, as a function of the
equilibrium energy/particle attained in the collisions described by the
parameter sets in Table~\protect\ref{tableA}.  The sharp increase in
$\rho_{l^+l^-}$ occurs after the plasma oscillation overpowers the viscous
collision term.}
\end{figure}

A further illustration of the effect is provided in Fig.~\ref{Ntaufig}.  The
equilibrium energy per particle, $\varepsilon/n$, grows with the violence of
the collision; i.e., with the value of $A_0$, as do the amplitude and
frequency of the plasma oscillations.  This effect is responsible for the
sharp increase in the dilepton yield, evident in Fig.~\ref{Ntaufig}, for the
most energetic collision in Table~\protect\ref{tableA}.  At lower values of
$\varepsilon/n$ the effect of plasma oscillations is suppressed by
collisions, but for the Set~4 initial conditions the magnitude and frequency
of the plasma oscillation are large enough to make their action evident in
spite of the damping, at least in our idealised treatment.  (NB.\ A rapid
expansion of the plasma will alter the initial conditions required for plasma
oscillations to have observable consequences.)

\section{Epilogue}
\label{sec:V}
We have used a quantum kinetic equation coupled with Maxwell's equation to
explore the formation and equilibration of a strong field plasma.  Using a
simple impulse model for the URHIC, which produces RHIC-like conditions, we
find that the non-Markovian aspects of the source term do not generate
observable effects.  However, the field-current feedback, which is
characteristic of the production of strongly coupled charges by a strong
field, manifests itself in the appearance of plasma oscillations in the
thermodynamic observables.  The oscillations are also evident in the
production rate of thermal dileptons and, while the time evolution of this
rate may not be measurable, the plasma oscillations act to significantly
enhance the time integrated rate.  The effect is marked by a sharp increase
in the dilepton yield when the energy per particle becomes large enough to
generate a high frequency and large amplitude plasma oscillation, which
initially overwhelms the effect of collisions.

While the magnitudes of the quantities we calculate are phenomenologically
reasonable, the primary results of our study are qualitative and many
improvements are possible.  Our relaxation time approximation to the
collision term is an intuitive and practical tool but a more realistic
connection with the actual collision process would provide a systematic and
well-constrained quantitative improvement.  A simpler step is the
introduction of a strongly momentum-dependent dressed-parton mass, which is
an essential feature of QCD~\cite{Mp2}.  That can have a significant impact
on the evolution of the plasma, promoting plasma oscillations~\cite{proc00},
and also on its subsequent hadronisation~\cite{huefner}.  Perhaps the most
significant defect of our study is the use of an Abelian model for the colour
fields and progress with a non-Abelian transport equation would be a marked
improvement~\cite{bahl2}.

\section*{Acknowledgments}
We thank R.\ Alkofer, V.\ Morozov and P.C.\ Tandy for helpful discussions.
This work was supported by: the Deutsche For\-schungs\-ge\-mein\-schaft under
project nos.\ SCHM~1342/3-1 and 436~RUS~17/102/00; the US Department of
Energy, Nuclear Physics Division, under contract no.\ W-31-109-ENG-38; the US
National Science Foundation under grant no.\ INT-9603385; and partly by the
Russian Federation's State Committee for Higher Education under grant no.\
N~97-0-6.1-4.  This research benefited from the resources of the National
Energy Research Scientific Computing Center.


\end{document}